\documentclass[aps,pre,preprint,groupedaddress]{revtex4}

\usepackage{graphicx}
\usepackage{latexsym}

\begin{document}


\title{
Direct observation of
the effective bending moduli of
a fluid membrane:
Free-energy cost due to the reference-plane deformations
}

\author{Yoshihiro Nishiyama}
\affiliation{
Department of Physics, Faculty of Science,
Okayama University, Okayama 700-8530, Japan.}


\date{\today}

\begin{abstract}
Effective bending moduli of a fluid membrane are investigated 
by means of the transfer-matrix method
developed in our preceding paper.
This method allows us to survey
various statistical measures for the partition sum.
The role of the statistical measures is arousing much attention,
since Pinnow and Helfrich claimed that under a suitable 
statistical measure, that is, the local mean curvature,
the fluid membranes are stiffened, rather than softened, 
by thermal undulations.
In this paper, 
we propose an efficient method
to observe the effective bending moduli directly:
We subjected a fluid membrane to a curved reference plane,
and from the free-energy cost due to the reference-plane deformations,
we
read off the effective bending moduli.
Accepting the mean-curvature measure, we found that
the effective bending rigidity gains even in the case
of very flexible membrane (small bare rigidity); 
it has been rather controversial that 
for such non-perturbative regime, the analytical prediction does apply.
We also incorporate the Gaussian-curvature modulus, and calculated
its effective rigidity.
Thereby, we found
that the effective Gaussian-curvature modulus 
stays almost scale-invariant.
All these features are contrasted with the results under
the normal-displacement measure.
\end{abstract}


\maketitle

\section{Introduction}
\label{section1}

Amphiphilic molecules in water segregate spontaneously into
flexible extended surfaces called fluid (lipid) membranes
\cite{Lipowsky91,Peliti96}.
The fluid membranes are free from both surface tension and shear modulus,
and the elasticity is governed only by bending rigidity
\cite{Canham70,Helfrich73}.
The Hamiltonian is given by the following form,
\begin{equation}
\label{Hamiltonian}
H= \int d A \left(
            \frac{\kappa}{2} J^2 + \bar{\kappa} K
           \right)   .
\end{equation}
The mean curvature $J$ is given by the summation of
two principal curvatures $J=c_1+c_2$, whereas the Gaussian curvature $K$ is
given by their product $K=c_1 c_2$.
The corresponding two moduli $\kappa$ and $\bar{\kappa}$ are
called bending rigidity and Gaussian-curvature modulus, respectively.
The integration $\int dA$ extends over the whole membrane surface.
The Gaussian-curvature term governs the global structure
of the membranes, because the term
$\int d A K$ measures a topological index; for instance,
for a vesicle with $n_h$ handles, 
such an identity
$\int d A K=4\pi (1-n_h)$ holds
(Gauss-Bonnet theorem).

In spite of its seemingly simple expression,
it is very hard to treat the Hamiltonian (\ref{Hamiltonian})
by analytical methods.
As a matter of fact,
when written in terms of an explicit parameterization,
the Hamiltonian
becomes ugly;
see Eqs. (\ref{Monges_J})-(\ref{Monges_dA}) mentioned afterwards.
Hence, owing to the thermal undulations and the mutual interactions,
it is expected that the effective bending moduli
are modified effectively for macroscopic length scales.
In order to clarify this issue,
numerous renormalization-group analyses have been reported so far
\cite{Peliti85,Forster86,Kleinert86,David89}.
For the bending rigidity, the following renormalization-group
equation has been obtained;
\begin{equation}
\label{renormalization_group_equation}
\kappa'=\kappa -\alpha \frac{k_B T}{8\pi} \ln M ,
\end{equation}
with renormalized bending rigidity $\kappa'$,
temperature $T$, and the number of decimated molecules $M$.
Literature agrees that the numerical prefactor in the above equation
is $\alpha=3$.
[A more detailed account of the historical overview
would be found in Ref. \cite{Helfrich98}.]
Because of $\alpha>0$, the effective bending rigidity is reduced by 
thermally activated undulations.
This conclusion might be convincing, because the membrane shape itself should
be disturbed by the thermal undulations.
As a matter of fact, it has been known that the
orientational correlation is lost for long distances \cite{deGennes82}.
It is quite natural to anticipate that membranes become flexible for
length scales exceeding this correlation length.

Recently, however, Pinnow and Helfrich \cite{Helfrich98,Pinnow00}
obtained a remarkable conclusion $\alpha=-1(<0)$.
The key ingredient of their new argument is that they considered the
role of measure factors for the partition sum.
They insist that the local mean curvature $J$ should be 
the right statistical measure rather than other measures 
such as the normal displacement
$h$ and the local tilt angle $\theta$.
[The normal displacement $h$ has been used as a standard
measure.
We will explain the $h$-based parameterization afterwards.]
After an elaborated calculation of the variable replacement
$h\to J$ and succeeding renormalization-group
analysis, the authors reach the conclusion of $\alpha=-1$.
Moreover, as for the Gaussian-curvature modulus,
they insist that the effective modulus $\bar{\kappa}'$ 
should remain scale-invariant;
\begin{equation}
\label{renormalization_group_equation2}
\bar{\kappa}'=\bar{\kappa} .
\end{equation}
This conclusion again contradicts the common belief
that $\bar{\kappa}$ would be enhanced for macroscopic length scales;
namely,
$\kappa'=\kappa + ( {5 k_B T}/{6\pi} ) \ln M$
\cite{Kleinert86,David89}.
This enhancement signals the topological instabilities.

The developments mentioned above all stem from
the approximation that the membrane is almost flat,
and the thermally excited undulations are extremely small.
In our preceding paper \cite{Nishiyama02},
we developed an {\it ab initio} simulation scheme  
in order to study the thermodynamics of a fluid membrane
beyond such perturbative level.
As a demonstration, we calculated the
transformation coefficient $d \kappa ' /d\kappa$,
which yields the 
direction of the renormalization-group flow 
through coarse-graining.
We found that the renormalization-group flow
is influenced significantly by the choice of the statistical measures.
In fact, under the mean-curvature measure, we observed an indication
that the effective bending rigidity flows toward the large-$\kappa$ direction.
We did not include the $\bar{\kappa}$ term in the preceding work.

In this paper, extending the preliminary analysis,
we propose an efficient method to observe the effective bending 
moduli directly:
We subject a fluid membrane to a curved reference plane.
From the free-energy cost due to the reference-plane deformations,
we read off the effective bending moduli.
Our data indicate definitely that under the mean-curvature measure,
the membrane stiffening occurs
even for the case of very flexible membrane $\kappa < 1$.
This result supports the picture that the stiffening,
contrary to our naive expectation, is driven by the thermal fluctuations.
We also calculate the effective Gaussian-curvature modulus,
and show that it remains almost scale-invariant.
Again, the result is
in good agreement with the Pinnow-Helfrich claim.

It has to be mentioned that the Monte-Carlo method
has been utilized successfully in the studies of membranes and 
vesicles \cite{Peliti96},
For the Monte-Carlo method, however, a tethered (polymerized)
membrane \cite{Kantor86} rather than a fluid membrane is more suited, 
because a membrane is implemented in a computer as an assembly
of molecules and junctions bearing close resemblance to a tethered
membrane.
(Note that because of the absence of shear modulus,
fluid membranes should have no internal structure.)
However, Gompper and Kroll succeeded in simulating fluid
membranes by the Monte-Carlo method, allowing reconstructions
of junctions during the simulation
\cite{Gompper00}.
They observed the topological instabilities 
with respect
to the variation of temperature and membrane concentration.
In fairness, it has to be mentioned that their Monte-Carlo
data indicate softening for lipid vesicles.

The rest of this paper is organized as follows.
In the next section, 
we describe our new approach to the direct observation
of the effective bending moduli.
We also explicate the outlines of the numerical method
which was reported previously \cite{Nishiyama02}.
In Sec. \ref{section3}, we present the numerical results.
We are mainly concerned in the case of the mean-curvature measure.
For a comparison, we will also present the data 
calculated under the 
normal-displacement measure.
With the first-principle simulation method, we will show that
the scenario advocated by Pinnow and Helfrich holds true
even for the 
non-perturbative regime.
In the last section, we give summary and discussions.

\section{
Legendre transformation and
the effective bending moduli
}
\label{section2}

In this section, we explain the idea for calculating
the effective bending moduli.
Mathematical formulas necessary in the succeeding numerical
simulations are derived.
We start with recalling the outlines of the
transfer-matrix  method proposed previously \cite{Nishiyama02}.
It is important to recognize the outlines 
of the transfer-matrix construction, because it
elucidates the underlying physical
idea of our effective-bending-moduli calculation.
In short, it has to be recognized that our membrane should be
classified into the ``open framed membrane'' in the category of 
Ref. \cite{Peliti96}.

\subsection{Transfer-matrix formalism: 
A brief remainder of Ref. \cite{Nishiyama02}}
\label{section2_1}

As mentioned in Introduction, the fluid membranes 
are free from shear modulus.
This fact tells that the fluid membranes should have no internal structure.
Hence, it is by no means fruitful to think of the microscopic
constituents realizing the Hamiltonian (\ref{Hamiltonian}).
Hence, we proposed in Ref. \cite{Nishiyama02}, an alternative,
in a sense, rather simple-minded, approach to the fluid membrane:
We constructed the transfer matrix directly from the Hamiltonian
(\ref{Hamiltonian}).
In the construction, we managed several discretizations which we explain
below.

As noted in Introduction, the membrane shape is parameterized
by the
normal (transverse) displacement $h(x,y)$ from a base (reference) plane.
The variables $x$ and $y$ denote the
Cartesian coordinates on the reference plane.
In term of this displacement field $h(x,y)$, the mean curvature and
the Gaussian curvature are parameterized explicitly as follows
\cite{Chaikin95};
\begin{equation}
\label{Monges_J}
J(x,y)=
\frac{
(\partial_x^2 h +\partial_y^2 h)(1+(\partial_x h)^2+(\partial_y h)^2)
-2\partial_x h\partial_y h\partial_x\partial_y h
-\partial_x^2 h(\partial_x h)^2
-\partial_y^2 h(\partial_y h)^2
}{
(1+(\partial_x h)^2 + (\partial_y h)^2)^{3/2}
}  ,
\end{equation}
and, 
\begin{equation}
\label{Monges_K}
K(x,y)=
\frac{
\partial_x^2 h \partial_y^2 h - (\partial_x \partial_y h)^2
     }{
(1+(\partial_x h)^2 + (\partial_y h)^2)^2
     }         ,
\end{equation}
respectively.
Similarly, the infinitesimal area $dA$ is given by,
\begin{equation}
\label{Monges_dA}
d A=(1+(\partial_x h)^2 + (\partial_y h)^2)^{1/2} dx dy .
\end{equation}
Putting them together into the Hamiltonian (\ref{Hamiltonian}), we arrive
at the explicit representation in terms of the displacement field $h(x,y)$.
Now, we are led to a two-dimensional scalar-field theory 
with considerably complicated interactions.
It is notable
that the theory has even no obvious perturbation parameter.
This fact motivated us to develop a first-principle simulation technique.

\begin{figure}
\includegraphics{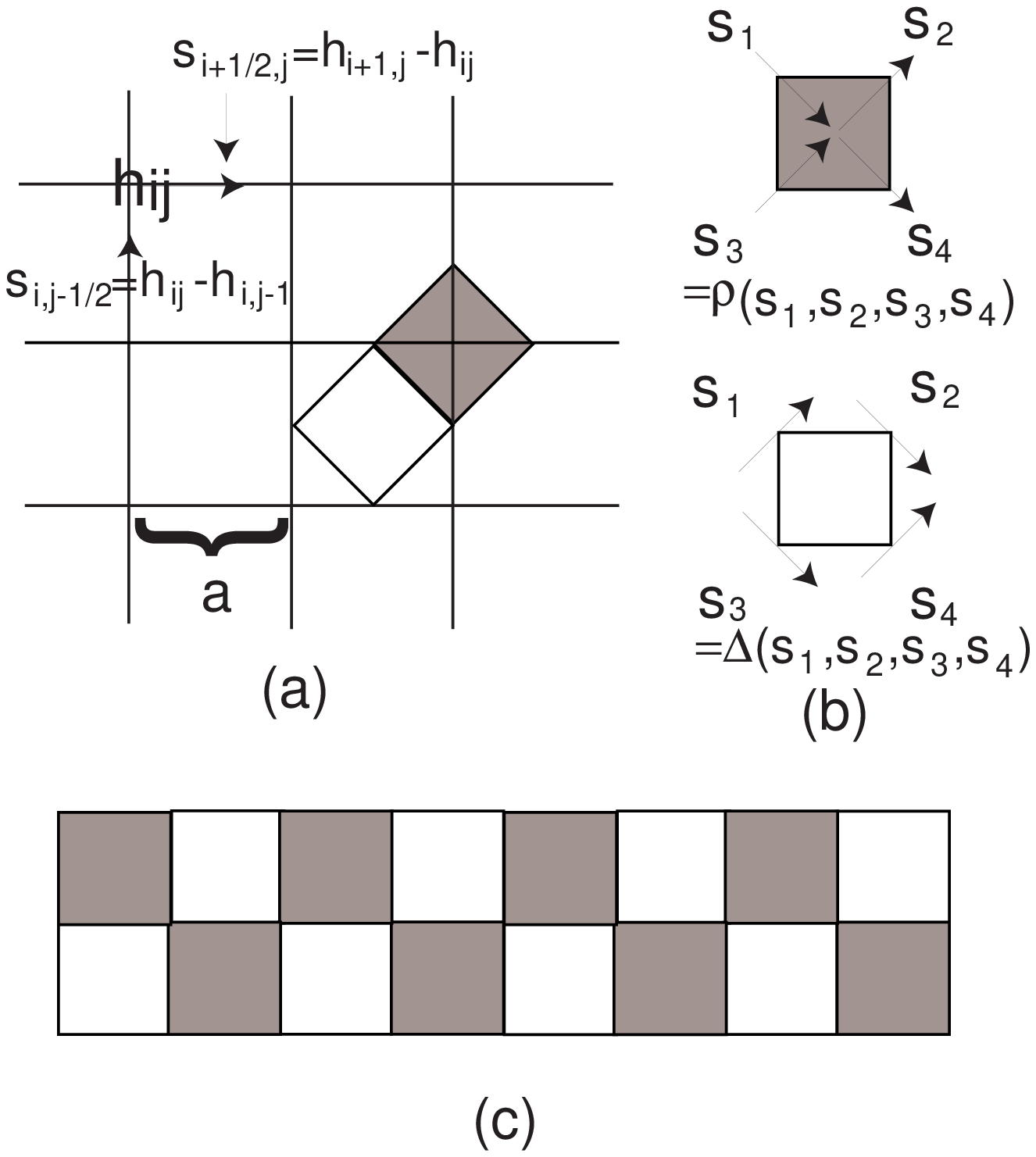}%
\caption{\label{figure1}
(a) On the square lattice,
we consider scalar field $h_{ij}$ denoting normal
displacement of a membrane with respect to a reference plane.
Step variable (gradient field) $\vec{s}=a\vec{\partial}h$ is defined
at each link.
(b) The local statistical weights $\rho$ 
[Eq. (\ref{statistical_weight1})]
and $\Delta$
[Eq. (\ref{statistical_weight2})]
are represented by shaded
and open squares, respectively.
The statistical weight $\rho$ has a variant so as to take account of
other integration measure such as the local mean curvature
(\ref{statistical_weight3}).
(c) From these local statistical weights, we construct
a strip whose
row-to-row statistical weight yields the transfer-matrix element.
This transfer matrix is diagonalized \cite{Nishiyama02} with the
DMRG method \cite{White92,White93,Nishino95}.
}
\end{figure}

We put the theory on a square lattice with lattice constant $a$; see 
Fig. \ref{figure1} (a).
Accordingly, the field variables are now indexed by integer indices; 
namely, $h(x,y) \to h_{ij}$.
Hereafter, we set the lattice constant as the unit of length;
namely, $a=1$.
Our theory has the translational invariance of $h \rightarrow h+\Delta h$,
and thus, the absolute value of $h$ is meaningless.
Therefore, it is sensible to use the link variable 
$\vec{s}=a\vec{\partial}h$ rather than $h$;
note that $\vec{s}$ denotes the step associated at each link;
see Fig. \ref{figure1}.
We are led to the dual lattice model.
Thereby,
for each shaded plaquette of this dual lattice (originally a vertex spanned
by four adjacent links), the following local statistical weight
is associated;
\begin{equation}
\label{statistical_weight1}
\rho(s_1,s_2,s_3,s_4)  
     =   \exp (  - {\cal H} )       ,
\end{equation}
with,
\begin{equation}
{\cal H}  = 
     d A(s_1,s_2,s_3,s_4)
     (\frac{\kappa}{2}
           J(s_1,s_2,s_3,s_4)^2
       +\bar{\kappa} K(s_1,s_2,s_3,s_4) )
                                ,
\label{local_Hamiltonian}
\end{equation}
where $J(s_1,s_2,s_3,s_4)$, $K(s_1,s_2,s_3,s_4)$ and 
$dA(s_1,s_2,s_3,s_4)$ are to be replaced \cite{Nishiyama02}
with the finite-difference versions of the differential forms
(\ref{Monges_J})-(\ref{Monges_dA}).
See Fig. \ref{figure1} (b) for the definitions of $\{s_\alpha\}$.
We have set $k_B T=1$, because this factor can be absorbed into the
bending moduli $\kappa$ and $\bar{\kappa}$.
Moreover, we should introduce yet another ``statistical weight''
for each open plaquette so as to impose 
the constraint of ${\rm rot}\vec{s}=0$; 
the gradient field should be rotationless.
That is,
\begin{equation}
\label{statistical_weight2}
\Delta(s_1,s_2,s_3,s_4) = \delta_{s_1+s_2-s_3-s_4,0}        .
\end{equation}

To summarize, we are led to the dual lattice model
\cite{Villain75}
with the step variable $\vec{s}$. There are two types of statistical weights
$\rho(s_1,s_2,s_3,s_4)$ (\ref{statistical_weight1}) and 
$\Delta(s_1,s_2,s_3,s_4)$ (\ref{statistical_weight2}),
which are arranged in the checkerboard
pattern.
Likewise the transfer matrix is constructed as a strip-like segment shown 
in Fig. \ref{figure1} (c).
It is a good position to mention a number of remarks:
First, the above theory takes the displacement variable $h_{ij}$
as the statistical measure.
As noted in Introduction, the mean-curvature statistical measure
is considered to be physically sensible.
The conversion of the statistical measure is
achieved 
by the redefinition of the statistical weight; namely,
\begin{equation}
\label{statistical_weight3}
\rho(s_1,s_2,s_3,s_4) \to \rho(s_1,s_2,s_3,s_4)
\sqrt{
\prod_{\alpha=1}^{4}
    \left|
\frac{\partial J(s_1,s_2,s_3,s_4)}{\partial s_\alpha}
    \right|
}            .
\end{equation}
The square root is intended to take the geometrical mean,
because each step variable $s_\alpha$ is sheared by an adjacent
plaquette as well.
Second,
the step variable is discretized as
$s_i=\delta_s (i -N_s/2-0.5)$ with $i = 1, \cdots ,N_s$. 
The unit of step $\delta_s$ is determined self-consistently during the
simulation by $\delta_s=R \sqrt{\langle s_i^2 \rangle}$.
This step-variable discretization is an influential factor concerning the 
reliability
of the present simulation, and its performance 
was checked previously  \cite{Nishiyama02}.
Third, in order to diagonalize the transfer matrix,
we utilized the
the density-matrix renormalization
group \cite{White92,White93,Nishino95,Peschel99}.
The method was invented, originally,
so as to investigate the highly-correlated systems
such as the Hubbard models and the spin chains.
Later on, it was extended to the field of soft materials
such as the lattice vibrations \cite{Caron96,Zhang98,Weisse00}, 
the quantum string \cite{Nishiyama01a,Nishiyama02a},
and the bosonic systems \cite{Nishiyama99,Nishiyama01b}.
We repeat density-matrix renormalization one after
another so as to reach sufficiently long transfer-matrix strip length
\cite{Nishiyama02}.
The number of states retained for a renormalized block
is an important technical parameter which is denoted by $m$.

One may wonder that the strong deformations increase the area of the piece
of membrane considered to such an extent that the emerging additional
degrees of freedom for additional molecules cannot be ignored.
According to the idea of Helfrich, however,
the correct statistical measure is the
local mean curvature that has noting to do with the ``molecules.''
Therefore, our treatment is justified even for such strong deformations
as long as we accept his idea.
Strictly speaking, we are not dealing with such ``molecules'' at all;
rather, we had just discretized the real space so as to form a square network,
and the vertices are not to be regarded as molecules.
In fact, the fluid membrane should have neither internal structure nor
fixed connectivity,
and it is not fruitful to think of microscopic constituents,
and regard them as the degrees of freedom.

Finally, in the above, we have postulated the presence of a reference plane
from which all undulations are created,
and we imposed no restriction to the total area of
the membrane.
Hence, in terms of the category of Ref. \cite{Peliti96}, the membrane
is to be classified into ``open framed membrane;''
see Ref. \cite{Fournier01} as well. 
The free energy per unit area of the reference plane
is the naturally observable quantity.
This leads to the idea that the effective bending moduli
would be
measured from the increase of the stress energy due to
the reference-plane deformations.
We will pursue this idea in the next subsection.

\subsection{Legendre transformation and 
the effective bending moduli
$\kappa_{eff}^{(S,L)}$ and $\bar{\kappa}_{eff}^{(S,L)}$}
\label{section2_2}

In this subsection, we explain our approach to the effective bending
moduli.
There are two types of bending moduli such as  $\kappa$ and $\bar{\kappa}$; 
see Eq. (\ref{Hamiltonian}).
First, we explain the way to calculate the effective bending rigidity.
We introduce the following Hamiltonian with an additional term;
\begin{equation}
\label{Hamiltonian_with_source_J}
H_C=H - \sum_i C J_i = \sum_i ( \frac{\kappa J_i^2}{2} dA_i 
   +\bar{\kappa} K_i d A_i - C J_i )  .
\end{equation}
The index $i$ runs over the shaded plaquettes of Fig. \ref{figure1};
The quantities $J_i$, $K_i$, and $dA_i$ are 
the same as those in Eq. (\ref{local_Hamiltonian}),
but possessing the plaquette index $i$ now.
Note that the additional term is not lumped together with the factor $dA_i$,
because our aim is to calculate the stress energy with
respect to the reference plane rather than the membrane surface itself.
The additional term breaks the symmetry of the mean curvature
$J \leftrightarrow -J$ linearly.
In other words,
the membrane is forced to bend so as to possess a non-vanishing
spontaneous mean curvature.
Hence, with respect to the stress energy due to the 
reference-plane deformation,
we are able to observe the effective bending rigidity.
That is the basic idea of our approach.
Such idea was featured in an analytical treatment as well \cite{Helfrich98}.
By the way, our aim is to put forward this idea to an actual
computer simulation.

As is well-known,
the above idea is best formulated by the Legendre
transformation;
\begin{equation}
G(j) =  F(C) +C j        ,
\end{equation}
with $\partial F/\partial C=-j$.
Here,
$F(C)$ denotes the free energy of the Hamiltonian 
(\ref{Hamiltonian_with_source_J})
per unit cell (one shaded plaquette).
$G(j)$ is the desired Legendre-transformed free energy,
which is a function of the spontaneous mean curvature $j$.
Our concern is to obtain the effective bending rigidity,
\begin{equation}
\kappa_{eff}^{(S)} = \frac{\partial^2 G}{\partial j^2}     .
\end{equation}
With use of the well-known identity
$(\partial^2 G/\partial j^2)(\partial^2 F/\partial C^2)=-1$,
we obtain the expression for 
the effective bending rigidity;
\begin{equation}
\label{effective_bending_rigidity_S}
\kappa_{eff}^{(S)} = - 1 / \frac{\partial^2 F}{\partial C^2}    .
\end{equation}
Let us address a number of remarks:
First, the free energy $F$ is readily accessible by
the transfer-matrix calculation.
Hence, the above formula is suited to our computer simulation.
The remaining task is the numerical differentiation.
We had adopted ``Richardson's deferred approach to the limit''
algorithm explicated in the textbook \cite{NRF}.
Second, we started from the Hamiltonian 
(\ref{Hamiltonian_with_source_J}), which is defined on the reference
plane rather than the original fluctuating membrane surface.
Therefore, the effective bending rigidity corresponds
to the elastic modulus with respect to the reference-plane deformation.
That is precisely what we sought.

The expression of Eq. (\ref{effective_bending_rigidity_S}) 
may seem to be exceedingly formal.
In fact, it may be unclear how the interaction of the curvature with
the undulations is taken into account.
It is noteworthy that the Hamiltonian contains the symmetry breaking
term $C J_i$ and we have to evaluate the free energy in the presence of it.
Just like the two-dimensional Ising model with the external field that
has not yet been solved exactly, such
problem with the symmetry breaking term is far from being trivial
in itself, and
generally, it contains valuable informations such as the
interaction between the background curvature and the thermal undulations.
In fact, Eq. (\ref{effective_bending_rigidity_S}) states that the 
second derivative with respect to $C$
yields the effective rigidity.
That is the underlying idea behind the formal expression of Eq. 
(\ref{effective_bending_rigidity_S}).
We stress that such an additional term is readily tractable
by our simulation method,
and owing to this advantage, we are able to access the effective rigidity
in a quite straightforward manner.

\begin{figure}
\includegraphics{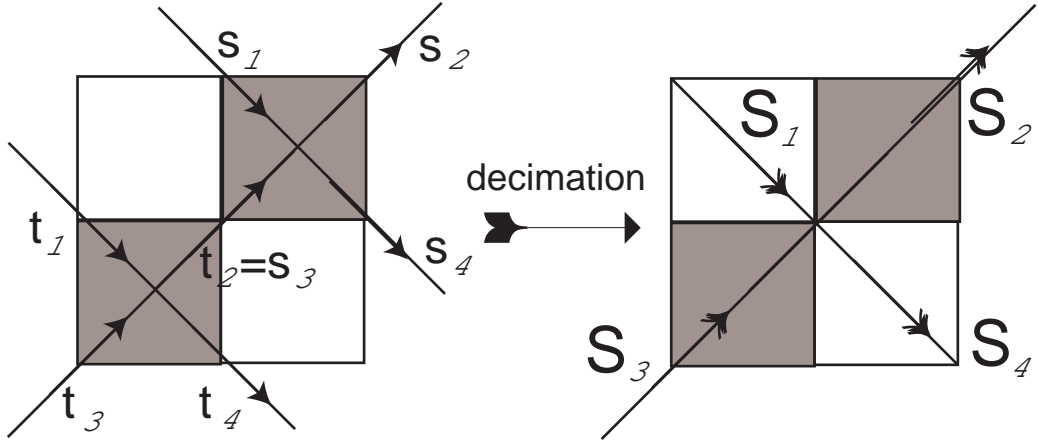}%
\caption{\label{figure2}
Real-space decimation procedure.
From the decimation,
coarse-grained curvatures $\tilde{J}$ 
and $\tilde{K}$ are constructed \cite{Nishiyama02}.
$\tilde{J}$ and $\tilde{K}$ are used so as to
obtain the corresponding effective bending moduli;
namely,
$\kappa_{eff}^{(L)}$ 
(\ref{effective_bending_rigidity_L}) and
$\bar{\kappa}_{eff}^{(L)}$
(\ref{effective_Gaussian_curvature_modulus_L}).
}
\end{figure}

We will introduce another effective bending rigidity:
Through the coarse-graining depicted in Fig. \ref{figure2},
we obtain a coarse-grained lattice
and the corresponding smeared curvature $\tilde{J}$ \cite{Nishiyama02}.
after rescaling the unit of length $\sqrt{2} a \to a$ and 
reexpressing the formulas in terms of the rescaled (smeared) quantities
[we put $\tilde{{}}$ for them],
we obtain the effective rigidity for the
coarse-grained membrane;
\begin{equation}
\label{effective_bending_rigidity_L}
\kappa_{eff}^{(L)}=-1/\frac{\partial^2 \tilde{F}(\tilde{C})}{\partial \tilde{C}^2} 
               .
\end{equation}
Note that two unit cells are renormalized into one
coarse-grained unit cell.
In other words, two molecules are renormalized into one decimated molecule,
and hence, the parameter $M$ in Eq. 
(\ref{renormalization_group_equation}) 
should be $M=2$.

Let us turn to the Gaussian-curvature modulus.
In this case, we incorporate the following additional term coupling 
to the Gaussian curvature linearly;
\begin{equation}
H_D = H - \sum_i D K_i =\sum_i ( \frac{\kappa J_i^2 }{2} dA_i + \bar{\kappa} K_i dA_i - D K_i )  .
\end{equation}
Similar to the above, this leads to the following Legendre transformation,
\begin{equation}
G(k)=F(D)+D k            ,
\end{equation}
with $\partial F/\partial D = -k$.
Our concern is to obtain the effective Gaussian-curvature modulus,
which, in other worlds, the effective symmetry breaking term with respect to
the $K$ field. 
Postulating that the effective free energy $G(k)$ is a quadratic polynomial
in terms of $k$,
we obtain the following expression,
\begin{equation}
\label{effective_Gaussian_curvature_modulus_S}   
\kappa_{eff}^{(S)}=
-
\frac{\partial F}{\partial D}  /  
  \frac{ \partial^2 F }{ \partial D^2 }
               .
\end{equation}
Again, the similar idea applies to the 
coarse-grained lattice.
Hence, we obtain,
\begin{equation}
\label{effective_Gaussian_curvature_modulus_L}
\kappa_{eff}^{(L)}=
-
\frac{\partial \tilde{F}}{\partial \tilde{D}}  /
  \frac{ \partial^2 \tilde{F} }{ \partial \tilde{D}^2 }
                           .
\end{equation}
We complete preparing the mathematical formulas
for the effective bending moduli.
We apply these formulas to the computer simulation in the next section.

\section{Numerical calculation of the effective bending moduli:
Role of the statistical measures}
\label{section3}

In this section, we explore the effective bending moduli
[$\kappa_{eff}^{(S,L)}$, Eqs. 
(\ref{effective_bending_rigidity_S}) and
(\ref{effective_bending_rigidity_L}); and
$\bar{\kappa}_{eff}^{(S,L)}$, Eqs.
(\ref{effective_Gaussian_curvature_modulus_S}) and
(\ref{effective_Gaussian_curvature_modulus_L})]
of
a fluid membrane by means of the transfer-matrix method
explained in Sec. \ref{section2_1}.
The transfer matrix is diagonalized \cite{Nishiyama02}
by means of 
the density-matrix renormalization group \cite{White92,White93,Nishino95}.
Making a comparison between the results under 
the mean-curvature 
and the normal-displacement measures,
we will elucidate the role of statistical measures.
As mentioned in the above section, 
we have fixed the temperature ($k_B T=1$),
because this factor can be absorbed into the redefinition
of the bending moduli.

Here, we shall outline some technical points, that are relevant to the
simulation precision.
[Detailed account of the simulation algorithm is presented in
Ref. \cite{Nishiyama02}.]
We repeated forty renormalizations for obtaining each plot;
namely, the strip length of the transfer matrix
extends to $L=80$.
The technical parameters $m$, $N_s$ and $R$ 
are indicated in each figure caption;
see Sec. \ref{section2_1} for
the meanings of these technical parameters.
These parameter values are equivalent or even improved
to those used previously \cite{Nishiyama02}.
Therefore, the reliability of the simulation is maintained.

\subsection{Effective bending rigidity $\kappa_{eff}^{(S,L)}$}

In this subsection, we focus our attention on the effective
bending rigidity.
For that purpose, for the time being, we drop the 
Gaussian-curvature term, which is considered in the next subsection;
namely, we set $\bar{\kappa}=0$.
In Fig. \ref{figure3}, we plotted the effective bending rigidity
$\kappa_{eff}^{(S,L)}$
for various bare rigidity $\kappa$.
The moduli
$\kappa_{eff}^{(S)}$ and $\kappa_{eff}^{(L)}$ denote
the effective bending rigidities for smaller and longer length scales,
respectively; see Eqs. 
(\ref{effective_bending_rigidity_S}) and
(\ref{effective_bending_rigidity_L}).
Here, we have accepted the local mean curvature $J$ as for the 
statistical measure;
recent through discussion \cite{Helfrich98,Pinnow00}
insists that this statistical measure should be the right one.
From the plot,
we see that the rigidities
$\kappa_{eff}^{(S)}$ and $\kappa_{eff}^{(L)}$
deviate from each other as for small $\kappa$.
In the small-$\kappa$ regime, the membrane becomes very flexible,
and so the thermal fluctuations should be enhanced significantly.
Hence, we see that the correction to the effective
rigidity is actually induced by the thermal fluctuations.
Moreover, we notice $\kappa_{eff}^{(L)} > \kappa_{eff}^{(S)}$.
Hence,
the membrane acquires stiffness for longer length scales;
namely, the membrane stiffening sets in.
Such membrane stiffening was first
predicted by the analytical arguments \cite{Helfrich98,Pinnow00}.
However, for the small-$\kappa$ regime, the analytical arguments
are not fully justified,
because the arguements stem from the ``nearly flat approximation.''
On the other hand,
our first-principle simulation does not rely on any perturbative treatment.
In that sense, our data demonstrate very definitely that 
the membrane stiffening occurs withstanding the thermal disturbances.

Note that the effective rigidity $\kappa_{eff}^{(S)}$ is by no means
identical to the ``bare'' coupling constant $\kappa$.
The former is the bending elasticity with respect to the reference-plane
deformations, whereas the latter is the elastic constant
of the membrane surface itself.
Therefore, they need not coincide.
However, for sufficiently large $\kappa$, as is seen from Fig. \ref{figure3},
the curve tends to be parallel to the slope of the line
$\kappa_{eff}^{(S)}=\kappa$,
indicating that they coincide asymptotically for large $\kappa$.

We shall argue the relationship between the above result
and our previous report \cite{Nishiyama02}.
The regime $\kappa < 0.4$, where we found a notable deviation of
$\kappa_{eff}^{(S)}$ and $\kappa_{eff}^{(L)}$
in Fig. \ref{figure3},
coincides with the area of the prominent  
$\partial \kappa' / \partial \kappa$ 
enhancement
reported in Fig. 7 of Ref. \cite{Nishiyama02}.
[Although 
$\partial \kappa' / \partial \kappa$ 
does not yield direct assessment of the effective rigidity,
we concluded that the $\partial \kappa' / \partial \kappa$ 
enhancement should reflect the membrane stiffening.]
Hence, in retrospect, our preceding analysis
appears to
capture the precursor of the membrane stiffening fairly correctly.

In Fig. \ref{figure3}, at $\kappa \approx 0.4$, there appears a singularity:
The moduli $\kappa_{eff}^{(S)}$ and $\kappa_{eff}^{(L)}$
approach to each other, and for
$\kappa > 0.4$, they split off again.
This singularity may indicate an onset of a phase transition.
For $\kappa > 0.4$, because of
$\kappa_{eff}^{(L)} > \kappa_{eff}^{(S)}$,
a membrane stiffening should occur.
As a matter of fact,
because of the discretization of the step variables 
[see Sec. \ref{section2_1}], 
it is likely that the membrane is trapped by the
most stable configuration (flat surface)
for large $\kappa$.
[It is expected that 
the membrane becomes flat
just like the smooth phase in the solid-on-solid model
with large surface tension.]
Moreover, it is well known that
the correlation length diverges exponentially for large $\kappa$
\cite{deGennes82}.
Such long correlation length would exceed the capability
of the numerical simulation.
The large-$\kappa$ behavior appearing in $\kappa>0.4$ is thus 
an artifact of the numerical simulation.
The membrane stiffening for large $\kappa$ is not intrinsic, and
is rather driven by 
the mechanism different from that 
of the small-$\kappa$ side.

In Fig. \ref{figure4} (a), keeping such drawback in mind, 
we have drawn an anticipated behavior of the effective 
bending rigidity for a wide range of $\kappa$.
As mentioned above,
the analytical treatment is justified for large $\kappa$.
On the other hand, our first-principle simulation 
works efficiently in the other side (non-perturbative regime).
For sufficiently large $\kappa$,
the analytical argument predicts the correction to the effective $\kappa$
such as
$\kappa'-\kappa= ln M /8 \pi=0.027 \dots$.
The correction appears to be exceedingly small
to be resolved by the numerical simulation; see Fig. \ref{figure3} as well.
On the other hand, for the small-$\kappa$ regime,
our simulation data indicate that the correction to the effective $\kappa$ 
increases very significantly.
The amount of correction is comparable to the thermal-fluctuation energy;
note that we have chosen $k_B T$ as the unit of energy ($k_B T=1$).
Hence, it is suggested that for macroscopic length scales,
the membrane acquires a considerable amount of effective stiffness, and
it would look almost flat irrespective of the thermal
disturbances.

Let us turn to the normal-displacement statistical measure.
In Fig. \ref{figure5}, we plotted
the effective bending rigidity $\kappa_{eff}^{(S,L)}$
under this statistical measure.
Notably enough,
the behavior is quite contrastive with that of the local-curvature measure
mentioned above:
The larger-scale effective rigidity $\kappa_{eff}^{(L)}$ is suppressed
by the thermal undulations.
Hence, it is shown that the membrane is softened effectively
for macroscopic length scales.
This result may meet our intuition, and has been predicted
by numerous analytical arguments based on the normal-displacement measure
\cite{Peliti85,Forster86,Kleinert86,David89}.
We stress that our first-principle simulation
covers various statistical measures in a unified way.
Our simulation clarifies fairly definitely that 
the choice of measure factors is vital for the thermodynamics
of the fluid membrane.
Again, for large $\kappa \approx 0.8$,
a signature
of the membrane stiffening $\kappa_{eff}^{(L)}>\kappa_{eff}^{(S)}$
comes up, and the membrane should undergo the flat phase.
This behavior is a drawback of our simulation as mentioned above.
Keeping this in mind,
we have drawn a schematic behavior of the effective bending rigidity
in Fig. \ref{figure4} (b).
For large $\kappa$, the analytical argument predicts 
the renormalization correction
$\kappa'-\kappa = -3 ln M/8\pi=-0.082\dots$, 
which is beyond the resolution of
the present numerical simulation.
As for the small-$\kappa$ regime, the correction 
is enhanced.
However, the enhancement is not so prominent as in the case of the
mean-curvature measure.
It is almost comparable to the analytical prediction;
see Fig. \ref{figure5} as well.
Hence, our first-principle simulation indicates that the
analytical formula (\ref{renormalization_group_equation}) 
is more or less applicable even for the
case of the non-perturbative regime under the 
normal-displacement measure.

\subsection{Effective Gaussian-curvature modulus $\bar{\kappa}_{eff}^{(S,L)}$}

In the above, we have studied the thermal-fluctuation-induced
corrections to the bending rigidity $\kappa$.
In this subsection, we incorporate the Gaussian-curvature modulus 
$\bar{\kappa}$,
and look into its effective strength $\bar{\kappa}_{eff}^{(S,L)}$.
As noted in Introduction, the Gaussian-curvature-modulus term
is related to the topological index, and hence, it
governs the global structure of the membranes.
Roughly speaking, for $\bar{\kappa} >0$, the plumber's-nightmare phase 
(lamellar with tunnel-like defects) is stabilized,
whereas for $\bar{\kappa} < 0$, the formation of vesicles (droplets)
is favored.
In that sense, the quantity $\bar{\kappa}_{eff}^{(S,L)}$ 
reflects the tendencies toward the topological instabilities.

In Figs. \ref{figure6} and \ref{figure7}, 
we presented the effective Gaussian-curvature modulus
for various bare $\bar{\kappa}$ under the statistical measures
of the mean curvature and the normal displacement, respectively. 
Here, $\kappa$ is fixed to be $\kappa=2/\sqrt{2}(=0.35\dots)$ and $0.4$ 
for respective figures.
First, let us argue the latter.
[Because this case provides a prototypical example, 
we will argue it prior to the mean-curvature case.]
The latter case has been studied extensively so far 
with analytical approaches \cite{Kleinert86,David89}.
However, because we are supposing that the membrane is framed
by the reference plane, there emerge some characteristic features.
For $\bar{\kappa} \approx 0$, we see that  
a large amount of effective $| \bar{\kappa}_{eff}^{(S)} |$ appears.
This result indicates that the membrane undulations
are dominated by the dimple-like deformations,
and possibly, the membrane tends to form droplets.
This feature is in accordant with the previous claim
\cite{Gompper98} that for small membrane concentration,
the membranes are thermodynamically unstable to the 
dissolution into the solvents (sponge phase).
On the contrary,
the effective modulus for the longer length scale, namely, 
$| \kappa_{eff}^{(L)} | $, exhibits considerable suppression.
Notably enough, it becomes even positive for $\bar{\kappa}>0.1$:
For macroscopic length scales,
the membrane recovers its stability around $\bar{\kappa} \approx 0$, 
although microscopic undulations are in favor of droplets.
Such crossover behavior is convincing, because the membrane is
framed by the reference plane, and macroscopically,
lamellar-type structure should be retained.

As the bare modulus $\bar{\kappa}$ decreases,
the effective moduli of different
length scales coincide at $\bar{\kappa} \sim -0.6$.
For the region exceeding this point $\bar{\kappa} < -0.6$, in turn,
$\bar{\kappa}_{eff}^{(L)}$ dominates $\bar{\kappa}_{eff}^{(S)}$ eventually.
Hence, in this region, 
the droplet formation is favored
for macroscopic length scales.
The location of this transition point is reminiscent of that 
advocated by the analytical arguments \cite{Golubovic94,Morse94},
which
predicts the transition point 
$\bar{\kappa}_c \approx -10\kappa/9=-0.44\dots$.
In our simulation, the planar-type morphology is
assumed a priori.
Therefore, in such regime $\bar{\kappa}<  \bar{\kappa}_c$,
our numerical simulation does not cover such sponge phase.
In fact,
it suffers from pathologies 
such as the diverging mean deviation of the step variables.
In the same way, for exceedingly large $\bar{\kappa} > 0.1$, 
the membrane becomes unstable owing to
the topological instability toward the plumber's-nightmare phase.
In the regions depicted in Figs. \ref{figure6} and \ref{figure7},
the planar-type morphology is retained,
and thus our simulations are reliable.

Second, 
let us turn to the case of the mean-curvature measure:
In Fig. \ref{figure6}, we plotted $\bar{\kappa}_{eff}^{(S,L)}$
for various bare Gaussian-curvature modulus $\bar{\kappa}$.
There appear some behaviors characteristic of the mean curvature measure:
We notice that 
$\bar{\kappa}_{eff}^{(S)}$ and 
$\bar{\kappa}_{eff}^{(L)}$ almost overlap each other around $\bar{\kappa}\approx0$.
That is, the Gaussian-curvature modulus stays
almost scale-invariant through coarse-graining.
This result supports the aforementioned analytical prediction 
\cite{Helfrich98,Pinnow00} of 
Eq. (\ref{renormalization_group_equation2}).
In addition to this, it is to be noted that $\bar{\kappa}_{eff}^{(S,L)}$
exhibits a large negative residual value $\bar{\kappa}_{eff}^{(S,L)} \approx -1.4$
around $\bar{\kappa}\approx0$.
That is,
although the bare coupling $\bar{\kappa}$ is turned off,
the membrane undulations are dominated by the dimple-like deformations.
This result validates the 
postulation by Helfrich \cite{Helfrich98} insisting that
the elementary excitations of the thermal undulations
should be the 
``hat excitations''
rather than the ordinary sinusoidal ones.
Because
the hat-excitation picture is the starting point of his arguments.
the whole theoretical theoretical structure appears to be selfconsistent
from our first-principle data.

\section{Summary and discussions}

\begin{figure}
\includegraphics{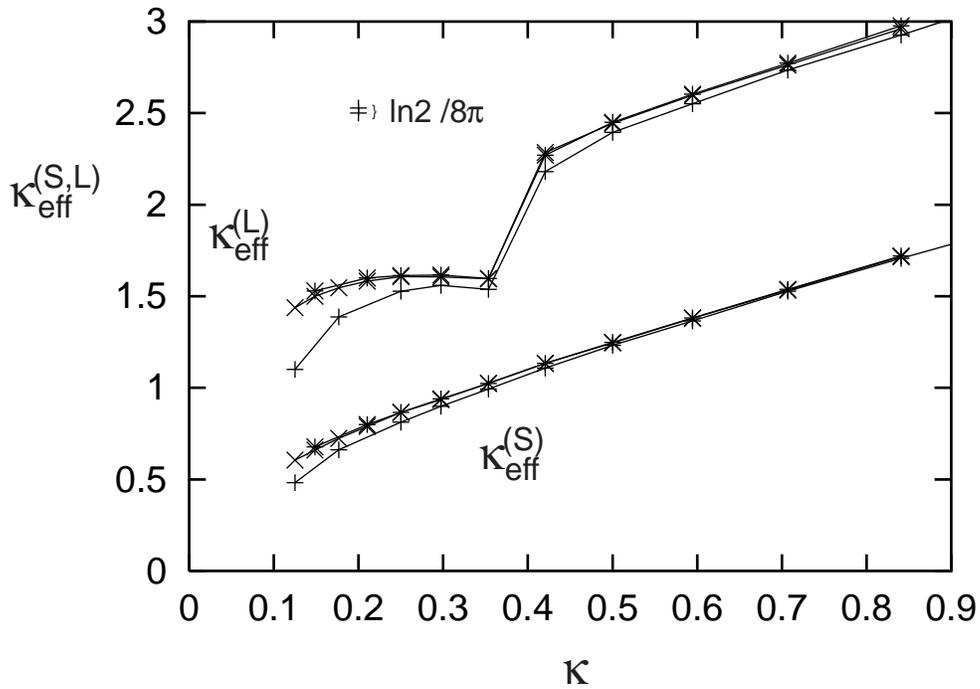}%
\caption{\label{figure3}
Effective bending rigidity $\kappa_{eff}^{(S,L)}$ is plotted for 
various bare bending rigidity $\kappa$ and 
the fixed Gaussian-curvature rigidity $\bar{\kappa}=0$.
We have accepted the local curvature as for the statistical measure.
The simulation parameters for each symbol are 
($+$) $m=15$, $N_s=8$, and $R=0.9$;
($\times$) $m=15$, $N_s=8$, and $R=0.8$;
and
($*$) $m=15$, $N_s=9$, and $R=0.7$.
Because of 
$\kappa_{eff}^{(L)} > \kappa_{eff}^{(S)}$, 
we see that the membrane is stiffened 
effectively for macroscopic length scales.
The stiffening in the large-$\kappa$ side, $\kappa>0.4$,
 may be the artifact of the numerical simulation
due to the pinning potential of discretized step variables;
there emerges the smooth phase just like the solid-on-solid model 
with large surface tension.
}
\end{figure}

\begin{figure}
\includegraphics{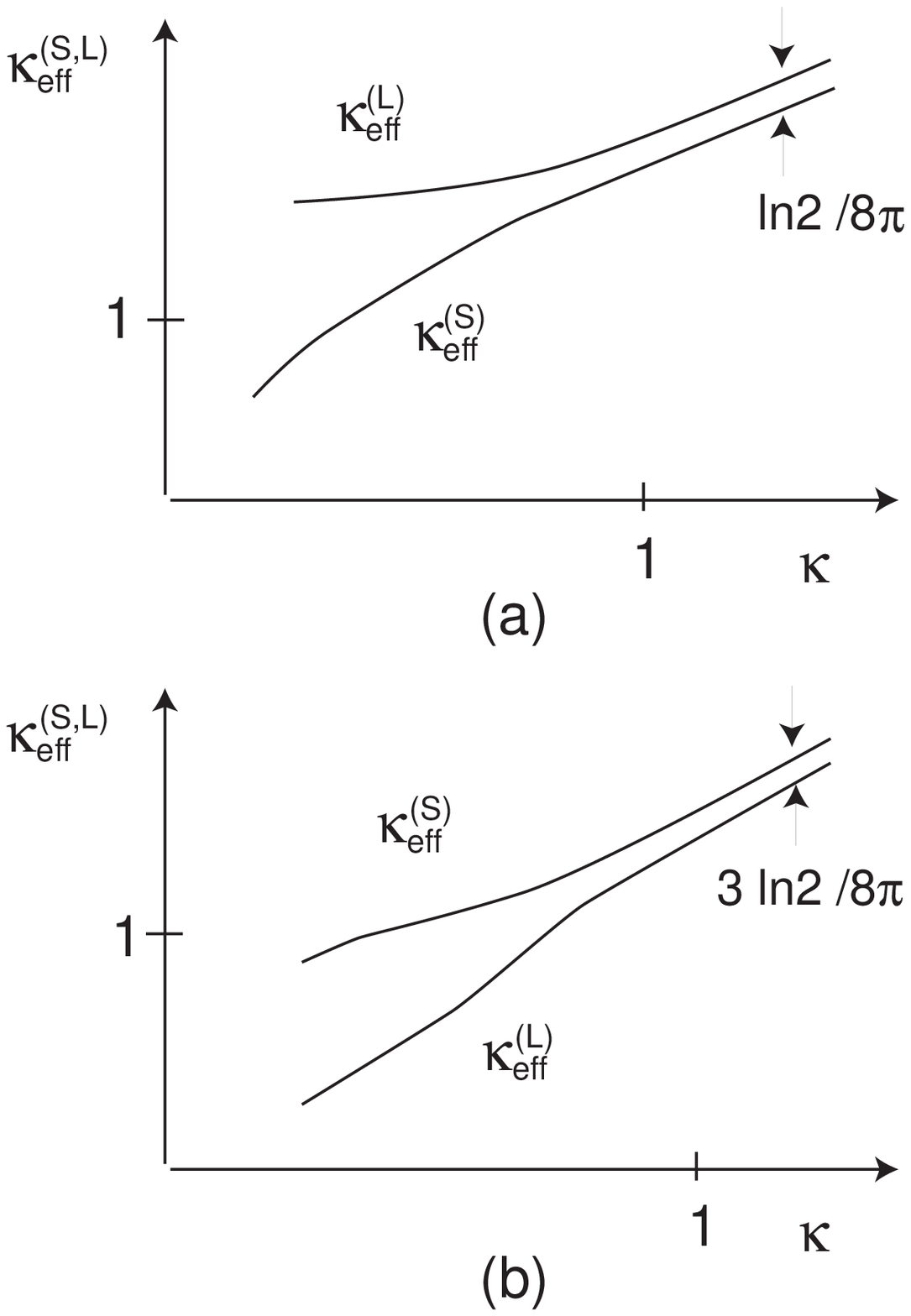}%
\caption{\label{figure4}
Schematic drawings of $\kappa_{eff}^{(S,L)}$ anticipated
from our first-principle data [Figs. \ref{figure3} and \ref{figure5}],
and the analytical result justified for sufficiently large $\kappa$
[Eq. (\ref{renormalization_group_equation})].
(a) The mean curvature is accepted as for the statistical measure.
(b) The normal displacement is accepted for the statistical measure.
}
\end{figure}

We have investigated the effective bending moduli,
$\kappa_{eff}^{(S,L)}$ [Eqs. 
(\ref{effective_bending_rigidity_S}) and 
(\ref{effective_bending_rigidity_L})] and 
$\bar{\kappa}_{eff}^{(S,L)}$ [Eqs. 
(\ref{effective_Gaussian_curvature_modulus_S}) and 
(\ref{effective_Gaussian_curvature_modulus_L})],
with an emphasis
on the role of the statistical measures for the partition sum.
We employed the transfer-matrix method developed in our preceding
paper \cite{Nishiyama02}, where we had reported a preliminary analysis
on the effective bending rigidity via the transformation coefficient
$\partial \kappa' / \partial \kappa$.
In the present paper,
we proposed the scheme to determine the effective bending moduli directly:
We calculated the free-energy cost due to the reference-plane deformations,
from which we read off the effective bending moduli. 
This idea is formulated in terms of the Legendre transformation,
and the mathematical formalism
is developed in Sec. \ref{section2_2}.
Based on the formulas, we carried out extensive computer simulations
in Sec. \ref{section3}.
As a result, we found a clear evidence of the membrane stiffening
in the case of the mean-curvature measure; see Figs. 
\ref{figure3} and \ref{figure4} (a).
The membrane stiffening was first predicted by the analytical approaches
\cite{Helfrich98,Pinnow00}, which are validated for sufficiently large $\kappa$.
Our first-principle data show that the membrane stiffening 
occurs even for the non-perturbative (small $\kappa$)
regime withstanding the thermal disturbances.
Surprisingly enough,
the enhancement of the effective bending rigidity
copes with 
the thermal-fluctuation energy $\sim k_B T$, 
suggesting that the membrane would stay almost flat 
for macroscopic length scales.

\begin{figure}
\includegraphics{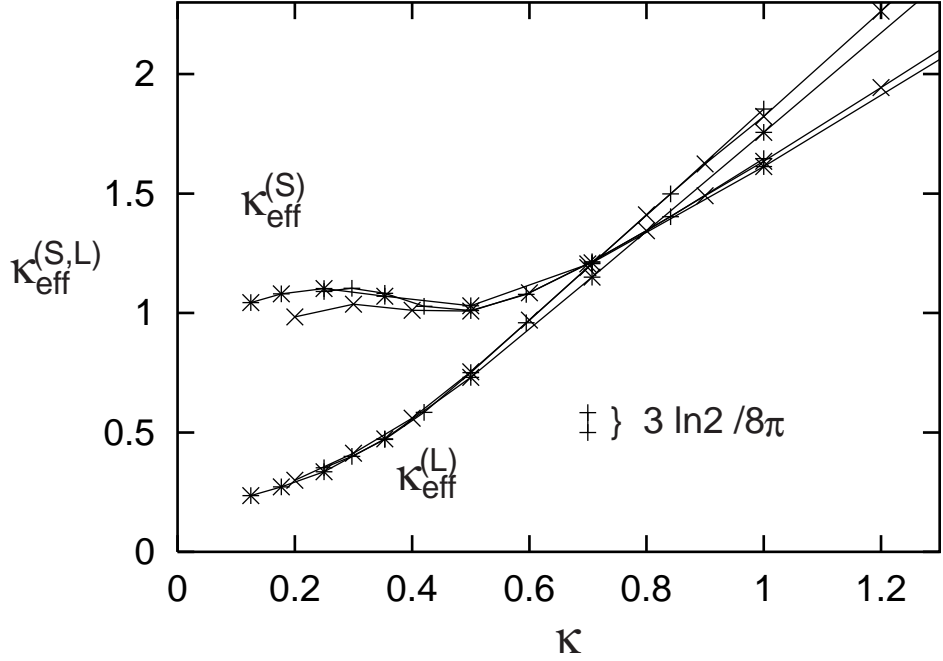}%
\caption{\label{figure5}
Effective bending rigidity $\kappa_{eff}^{(S,L)}$ is plotted for 
various bare bending rigidity $\kappa$ and
the fixed Gaussian-curvature rigidity $\bar{\kappa}=0$.
We have accepted the normal displacement as for the statistical measure.
The simulation parameters for each symbol are 
($+$)     $m=13$, $N_s=9$, and $R=0.55$;
($\times$)  $m=10$, $N_s=11$, and $R=0.45$; 
and
($*$)     $m=11$, $N_s=10$, and $R=0.5$.
Because of 
$\kappa_{eff}^{(L)} < \kappa_{eff}^{(S)}$, 
we see that the membrane is softened
effectively for macroscopic length scales.
The stiffening in the large-$\kappa$ side, $\kappa>0.8$,
may be the artifact of the numerical simulation
due to the pinning potential of discretized step variables;
there emerges the smooth phase just like the solid-on-solid model 
with large surface tension.
}
\end{figure}

On the contrary, under the 
normal-displacement statistical measure,
we found a clear indication of the membrane softening; see Figs. 
\ref{figure4} (b) and \ref{figure5}.
This fact indicates that the choice of measure factors
is indeed significant.
The correction to the effective bending rigidity
appears to be moderate compared with
that of the mean-curvature measure.
In fact, it is almost comparable to the
prediction by the analytical treatment of Eq. 
(\ref{renormalization_group_equation}) even for small $\kappa$.

For exceedingly large rigidity,
the membrane fluctuations freeze
because of the exponentially diverging correlation length
\cite{deGennes82}
and the pinning potential due to the step-variable discretization;
the membrane undergoes the flat phase eventually just like 
the solid-on-solid model with large surface tension.
The appearance of such phase is a drawback of the numerical simulation,
and in this respect, 
the simulation and the analytical treatment are both complementary.

\begin{figure}
\includegraphics{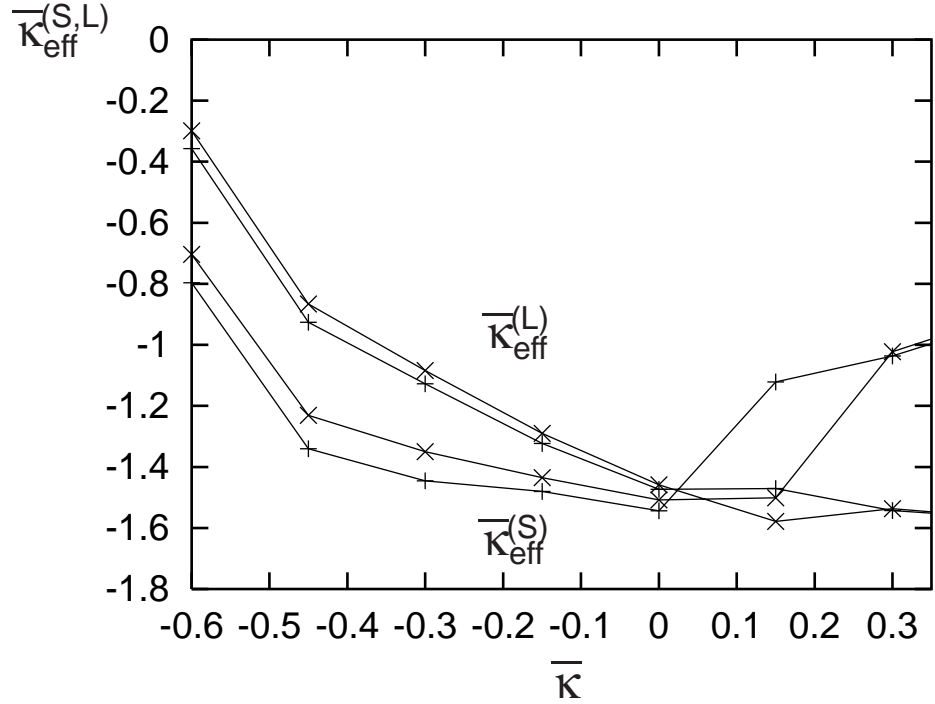}%
\caption{\label{figure6}
Effective Gaussian-curvature modulus $\bar{\kappa}_{eff}^{(S,L)}$ 
[Eqs. 
(\ref{effective_Gaussian_curvature_modulus_S}) and
(\ref{effective_Gaussian_curvature_modulus_L})]
is plotted for various bare modulus
$\bar{\kappa}$ and the fixed bending rigidity 
$\kappa =2/\sqrt{2}(=0.35\dots)$. 
We have accepted the local curvature as for the statistical measure.
The simulation parameters for each symbol are 
($+$) $m=15$, $N_s=7$, and $R=1$;
and
($\times$) $m=15$, $N_s=8$, and $R=0.9$.
We see that the effective modulus stays almost scale-invariant
around $\bar{\kappa}_{eff}^{(S,L)} \approx 0$ through coarse-graining, 
confirming the validity of the analytical prediction
[Eq. 
(\ref{renormalization_group_equation2})].
Moreover,
we notice that 
$\bar{\kappa}_{eff}^{(S,L)}$ exhibits a large negative residual value
even for zero bare $\bar{\kappa}=0$.
This fact 
reflects that the membrane undulations are dominated by
the formation of ``hat excitations'' \cite{Helfrich98}.
}
\end{figure}

\begin{figure}
\includegraphics{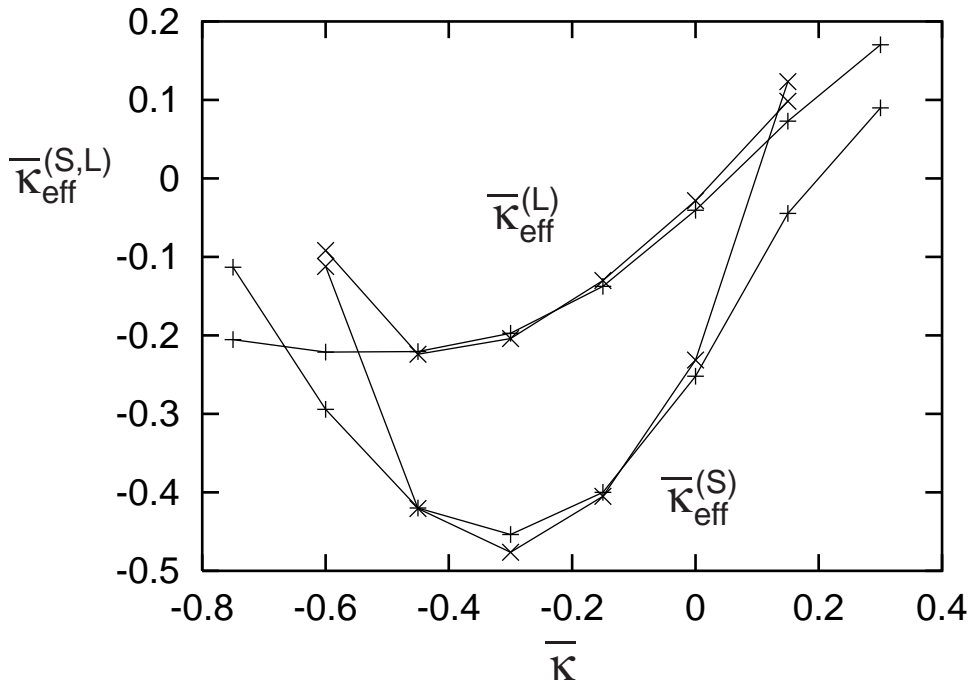}%
\caption{\label{figure7}
Effective Gaussian-curvature modulus $\bar{\kappa}_{eff}^{(S,L)}$ 
[Eqs. 
(\ref{effective_Gaussian_curvature_modulus_S}) and
(\ref{effective_Gaussian_curvature_modulus_L})]
is plotted for various bare modulus 
$\bar{\kappa}$ and the fixed bending rigidity $\kappa=0.4$. 
We have accepted the normal displacement as for the statistical measure.
The simulation parameters for each symbol are 
($+$) $m=14$, $N_s=8$, and $R=0.6$;
and 
($\times$) $m=9$, $N_s=11$, and $R=0.45$.
We see that $ | \bar{\kappa}_{eff}^{(L)} | $ is
suppressed around $\bar{\kappa} \approx 0$.
This fact tells that 
for macroscopic length scales
(because of the restriction of the reference plane)
the planar-type morphology is favored; namely,
$\bar{\kappa}$ is irrelevant in the infrared limit.
For $\bar{\kappa} < -0.6$, in turn, 
$ | \bar{\kappa}_{eff}^{(L)} | $ is enhanced eventually,
suggesting that the membrane tends to dissolve into the
solvent (droplet phase).
}
\end{figure}

Furthermore,
we incorporated the Gaussian-curvature modulus, and studied its effective
strength.
Accepting the mean-curvature measure, 
we found that the effective Gaussian-curvature moduli
for different length scales overlap each other around $\bar{\kappa} \approx 0$;
see Fig. \ref{figure6}.
In other words, the Gaussian-curvature modulus stays
almost scale-invariant through coarse-graining.
This fact is in good agreement with
the analytical prediction of Eq. 
(\ref{renormalization_group_equation2}).
In addition to this,
the effective Gaussian-curvature modulus exhibits a large negative residual
value even for zero bare modulus.
This fact indicates that the membrane fluctuations are 
governed by the dimple-like deformations that should be scale-free.
This observation again supports the scenario of Ref. \cite{Helfrich98}
insisting that the thermal undulations should be decomposed into
the hat excitations.

To conclude, by means of the first-principle simulation technique,
we have investigated the series of analytical predictions 
advocated by Pinnow and Helfrich \cite{Helfrich98,Pinnow00}.
Thereby, we found that these predictions hold true even for the
non-perturbative regime.
In particular,
the hat-excitation picture, which is the very starting point
of their argument, is validated by our  
$\bar{\kappa}_{eff}^{(S,L)}$ data.
Therefore, the postulation and its deductive hypotheses
turn out to be fairly consistent as a whole.
Hence, it is very likely that the mean curvature is indeed
a physically sensible 
statistical measure for the partition sum.

\begin{acknowledgments}
This work is supported by Grant-in-Aid for
Young Scientists
(No. 13740240) from Monbusho, Japan.
\end{acknowledgments}

\end{document}